\documentclass{PoS}
\usepackage[utf8]{inputenc}

\usepackage{booktabs,multirow}
\usepackage{array}
\usepackage{amsmath,amsfonts,amssymb,dsfont,mathtools,braket}
\usepackage{caption}

\title{Topology and glueballs in $SU(7)$ Yang-Mills with open boundary conditions}

\ShortTitle{Topology and glueballs in $SU(7)$ Yang-Mills with open boundary conditions}

\author{\speaker{Alessandro Amato}\\
       Helsinki Institute of Physics and University of Helsinki, P.O.Box 64, FIN-00014, Finland\\
       E-mail: \email{alessandro.amato@helsinki.fi}}

\author{Gunnar Bali\\
        Institute f\"ur Theoretische Physik, Universit\"at Regensburg, D-93040 Regensburg, Germany\\
        TIFR, Homi Bhabha Road, Mumbai 400005, India\\
        E-mail: \email{gunnar.bali@ur.de}}

\author{Biagio Lucini\\
        College of Science, Swansea University, Singleton Park, Swansea SA2 8PP, UK \\
        E-mail: \email{a b.lucini@swansea.ac.uk}}
\abstract{It is well known that the topology of gauge configurations generated in a Markov Monte-Carlo chain freezes as the continuum limit is approached. The corresponding autocorrelation time increases exponentially with the inverse lattice spacing, affecting the ergodicity of the simulation. In $SU(N)$ gauge theories for large N this problem sets in at much coarser lattice spacings than for $N=3$. This means that its systematics can be studied on lattices that are smaller in terms of the number of lattice sites. It has been shown that using open boundary conditions in time allows instantons to be created and destroyed, restoring topological mobility and ergodicity. However, with open boundary conditions translational invariance is lost and the influence of spurious states propagating from the boundary into the bulk on physical correlators needs to be carefully evaluated. Moreover, while the total topological charge can be changed, the mobility of instantons across the lattice is still reduced. We consider $SU(7)$ Yang-Mills theory and analyse its topological content in the periodic and open boundary condition cases. We also investigate scalar and pseudo-scalar glueball correlation functions.}

\FullConference{The 33rd International Symposium on Lattice Field Theory\\
		 14 -18 July  2015\\
		 Kobe International Conference Center, Kobe, Japan}

\begin{document}

\section{Introduction}
Lattice QCD has become a very reliable tool to produce first principles calculation. When the continuum limit is approached, a problem arises when periodic boundary conditions are  in use. In fact, the transition probability between different topological sectors of the gauge fields is suppressed towards large values of the lattice coupling parameter $\beta$ and the Markov chain produced in this way might lack of ergodicity. This therefore introduces a systematic error which is difficult to take into account and might lead to wrong results, at least for those observable which are sensitive to the topological content of the theory. In this work we will address this issue and in particular we will explore the use of open boundary conditions \cite{obc}, which are now starting to be used for a more systematical and well-defined study of glueball observables \cite{indobc} and for the meson and baryon spectrum \cite{obc-spectrum}. 

In this work, we focus on the $N=7$ Yang-Mills theory, relevant to the generalisation of QCD to the large-$N$ limit,  where the problem of frozen topology is worsened by a mechanism which suppresses the creation of small instantons. 
In order to better understand the origin and the consequences of this slowing down, we study the Monte Carlo evolution of the topological charge on periodic and open temporal boundary lattices.
Additionally, we investigate the instanton size distribution, which reveals the  presence of small dislocations near the boundary, crucial for global changes in topology. We also measure correlators of plaquettes and of the topological charge density in the two cases.

\section{Numerical simulations}

Simulations are performed at  one value of the gauge coupling $\beta =34.8343$, the same used  for meson spectroscopy studies on PBC lattices in~\cite{Bali:2013kia}. At this coupling, the square root of the string tension was found to be $a \sqrt{\sigma} \simeq 0.2093$, which gives $a \simeq 0.94$ fm, assuming $\sigma = (440\,\text{MeV})^2$. Configurations were saved every $200$ composite sweeps  with each composite sweep comprising of $1$ heatbath followed by $4$ overrelaxation updates.  
We present results for only one volume, which has $N_s=16$ points for the spatial directions and  $N_t=64$ for the temporal one.  We use lattices   that are either periodic in all directions (referred to as PBC lattices) or lattices  periodic in the three spatial directions and open in the temporal direction (OBC lattices). Both lattices have a statistics of $N_\text{CFG}\sim 3000$.
Using the same generation parameter for the OBC and the PBC case allows for a meaningful comparison of correlations and statistical errors in terms of the number of sweeps. In the
updates, the $SU(7)$ links were generated using a Cabibbo-Marinari procedure. A technical description of a simulation with open boundary conditions (OBC) can be found, e.g. in~\cite{obc}.

\section{Monte Carlo evolution of the topological charge}

We study the autocorrelation of the total topological charge $Q$. On a lattice, in terms of the plaquettes, this quantity is defined as 
\begin{eqnarray}
\label{eq:topcharge}
Q = \frac{1}{32 \pi^2}  \sum_{t= 0}^{N_t-1}q(t)\,, \qquad 
q(t) = \sum_{\vec x}  \sum_{i, \mu < \nu} \epsilon_{\mu \nu \rho \sigma} \,U_{\mu \nu}(t,\vec x) U_{\rho \sigma}(t,\vec x) \,,
\end{eqnarray}
where $q(t)$ is the topological charge density per time-slice.
Traditionally, one way to filter the ultraviolet fluctuations affecting this operator has been cooling~\cite{Teper:1985rb}. More recently,  another smoothing procedure, the Wilson flow~\cite{Luscher:2010iy}, has been proposed. The advantage of the Wilson flow over  cooling is that, in the former case, the smoothing procedure is continuous (i.e. it depends on a continuous parameter, the flow step, that can be taken infinitesimal) and is related to the asymptotic solution of classical equations that drive the system towards the minimum of the action. In~\cite{Bonati:2014tqa} it was shown that topological properties are insensitive to the chosen smoother and in our study we have used both methods and found compatible results. 
%

\begin{figure}[t]
	\center
	\includegraphics{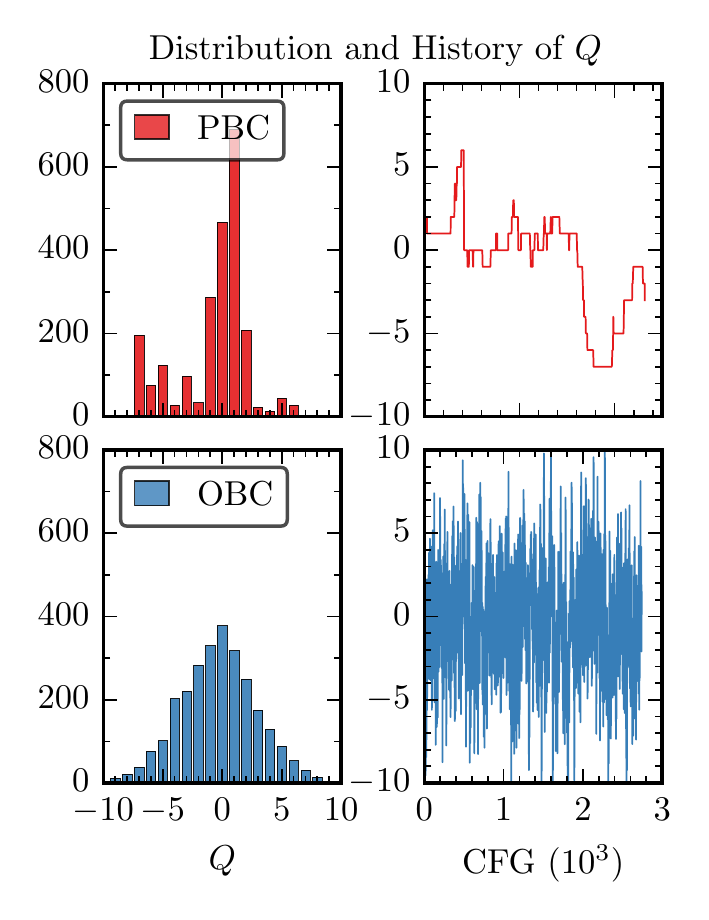}%
	  \includegraphics{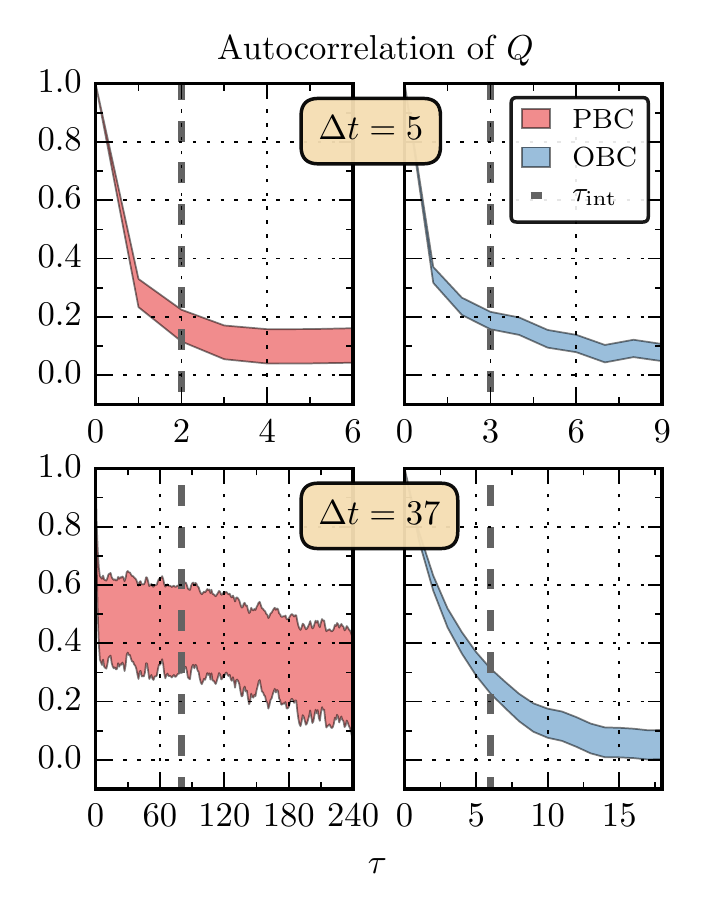}
	\caption{(Left) Distribution and Monte Carlo history of the topological charge on a $16^3 \times 64$ lattice. (Right) Monte Carlo time correlator of the topological charge on $16^3 \times 64$ lattices as a function of the number $\Delta t=5,37$ of time slices considered. The dashed line indicates the computed integrated autocorrelation time $\tau_\text{int}$. }
	\label{fig:q-dist}
\end{figure}

 We report the Monte Carlo history and    distribution of the total topological charge in Fig.~\ref{fig:q-dist}~(left) for both open and periodic boundary conditions. Smaller lattices provide a similar overall picture. The Monte Carlo history shows a wider fluctuating topological charge for the OBC case, which is an indication of good decorrelation. Conversely, topology on the PBC set takes much longer to change.
%
Following Ref.~\cite{Wolff:2003sm}, we compute\footnote{This computation needs to be taken with a pinch of salt when the temporal series has a length that is comparable to the correlation time, as it is the case for our largest lattice with PBC. However, here we are interested in estimates rather than in exact quantifications, and for our purposes the computation is adequate.} the integrated autocorrelation time $\tau_\text{int}$. 
We  consider temporal slabs of size $\Delta t$ centred at $(N_t +1)/2$  for the computation and in Fig.~\ref{fig:q-dist}~(right) we show the results for two values of $\Delta t$. 
Since the effect of boundaries is expected to be negligible at large $N_t$,  restricting the computation on a temporal slab of the lattice should give comparable results for both PBC and OBC, as shown by the case with $\Delta t=5$. As we include more temporal slabs in the computation,  $\tau_\text{int}$ increases significantly in the PBC case, while it remains small for the OBC  one.


\section{Instanton dynamics}

\begin{figure}[t]
\center
\includegraphics{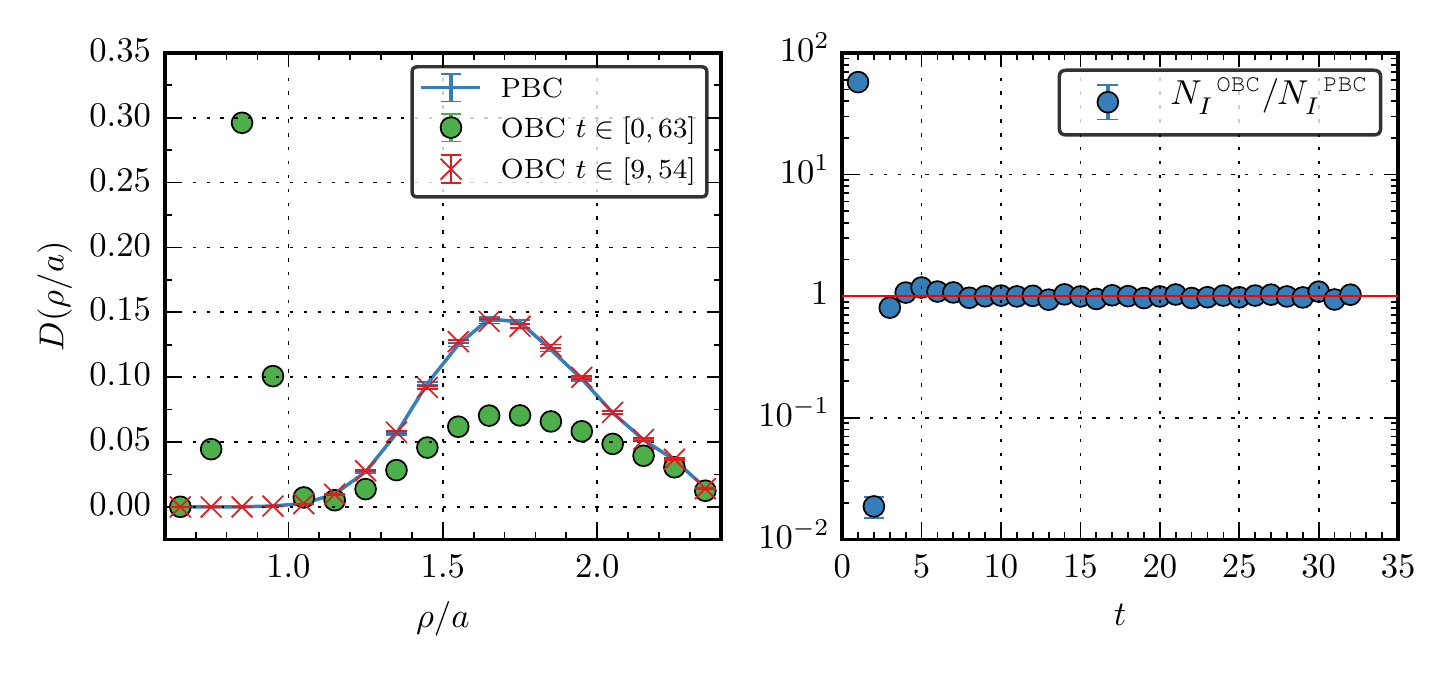}
\caption{Left: instanton size distribution on a  $16^3 \times 64$ lattice, on PBC, OBC including time slices close to the boundaries and OBC considering only slabs far from the boundaries. Right: ratio of the total number of instantons $N_I=N^++N^-$ with OBC over the same quantity with PBC on a $16^3 \times 64$ lattice as a function of the time coordinate.} 
\label{fig:instantons}
\end{figure}

The total topological charge $Q$  of a configuration can be viewed in terms of instantons  as $Q = N^{+} - N^{-}$, where $N^{+}$ ($N^{-}$) is the total number of instantons with positive (negative) topological charge (for a review see e.g.~\cite{Schafer:1996wv}).
In the instanton gas model, it is possible to carry out semiclassical calculations under the assumption that instantons are diluted and can be treated independently. This model, though too simplistic since it neglects interactions, proves to be adequate for the semi-quantitative level of our discussion. Identifying instantons on a lattice is not a straightforward process and here we follow~\cite{Smith:1998wt}.


A simple large-$N$ argument shows that the expected size distribution of small instantons, $D(\rho)$, where $\rho$ is the size in lattice units, scales as $\rho^{\frac{11}{3} N - 5} $ with the number $N$ of colors. This shows that small  instantons are suppressed at large $N$. On a lattice, an (anti-)instanton is created or destroyed at the ultraviolet length scale $a$. The appearance or disappearance of (anti-)instantons is the physical mechanism for the change of the topological charge in a Monte Carlo sequence. We expect only instantons at the typical physical length of the system to influence the dynamics. This means that, in order to produce a change, instantons either need to grow from the ultraviolet to the physical scale or shrink from the physical scale to the ultraviolet. The large-$N$ suppression of instantons at the ultraviolet scale is a plausible explanation for the observed slow change in topology with PBC as $N$ increases. This mechanism has already been studied in~\cite{Lucini:2001ej,Lucini:2004yh}. In our work, we wish to compare the instanton dynamics on periodic and open boundary lattices.

The size of instantons is determined from local peaks of the topological charge density using Eq.~(12) of ~\cite{Smith:1998wt}. The distribution in size (which we normalise to $1$) can then be worked out by analysing the whole Monte Carlo series. An analysis on our configurations gives the distributions reported in Fig.~\ref{fig:instantons} left. The distribution obtained with PBC is indistinguishable from the one obtained with OBC when the observations are restricted to a slab far from the boundaries, i.e. $9 \le t \le 54$. In particular, both cases show the expected suppression of small size instantons. When all temporal slices are considered for the OBC system, we observe a sharp peak emerging at small sizes. The implication of the result for the size distribution would be that small instantons localised near the open boundaries are crucial at changing the topology, while the open boundary and periodic systems have the same size distribution far in the bulk.

To corroborate this conclusion, we have studied the total number of instantons $N_I = N^+ + N^-$ as a function of the time slice. Fig.~\ref{fig:instantons} right reports the ratio between this quantity computed on the open boundary system ($N_I^{OBC}$) and the same observable computed with PBC ($N_I^{PBC}$). While the ratio becomes clearly one after 5 lattice spacings, its behaviour near the boundary indicates a surge of topological activity (which, from the size distribution, appears to be due to small dislocations) on the first slice followed by a sharp suppression on the second slice, while for $t > 2$, the ratio settles near $1$. 
A possible explanation is that the open boundary acts as a source and attractor of dislocations. Only few dislocations are able to penetrate the bulk, but they seem to be crucial for generating the right topological charge distribution.

\section{Glueball correlators and glueball masses}
The slow dynamics of the topological charge with PBC could in principle affect spectral masses, and in particular the masses of states that couple to the topological charge density $q(t)$. Since the latter quantity is negative under parity, for a SU($N$) gauge theory, one might expect the $0^{-+}$ glueball  mass to have a systematic error relating to the slow modes of $Q$. Moreover, the loss of ergodicity with PBC could also create a less direct systematic effect on other spectral masses. We  then investigate the plaquette-plaquette correlator (from which we can extract the mass of the $0^{++}$ glueball) and the correlator of $q(t)$ (whose asymptotic decay gives the mass of the $0^{-+}$ glueball) on periodic and open boundary lattices. In order to compare the spectral tower, we  study the ratio of those correlators at varying temporal separation $t$. 
The correlators $G_O(t)$ are defined as:
\begin{equation}
	G_O(t)=\frac{1}{n_{t_0}} \sum_{i=1}^{n_{t_0}}\braket{ O(t_0^i)\,O(t_0^i+t) }
\end{equation}
where  the sum is carried out over a set of $n_{t_0}$ time slices and the operator $O(t)$ is either $q(t)$ or the spatial plaquette averaged over the volume. 
In the case of PBC we use all available time slices, i.e. $n_{t_0}=N_t$ with $t_0= 0,\dots,N_t-1$  and the parameter $t$ runs from $0$ to $N_t/2+1$. In the case of OBC,  we chose $n_{t_0}=8$ values for $t_0$ far from the boundary, specifically  $t_0=15,\dots,18$ with $t=0,1,\dots,(N_t/2+1)$ and $t_0=45,\dots,48$ with $t=0,-1,\dots,-(N_t/2+1)$.

\begin{figure}[t]
	\center
        \includegraphics{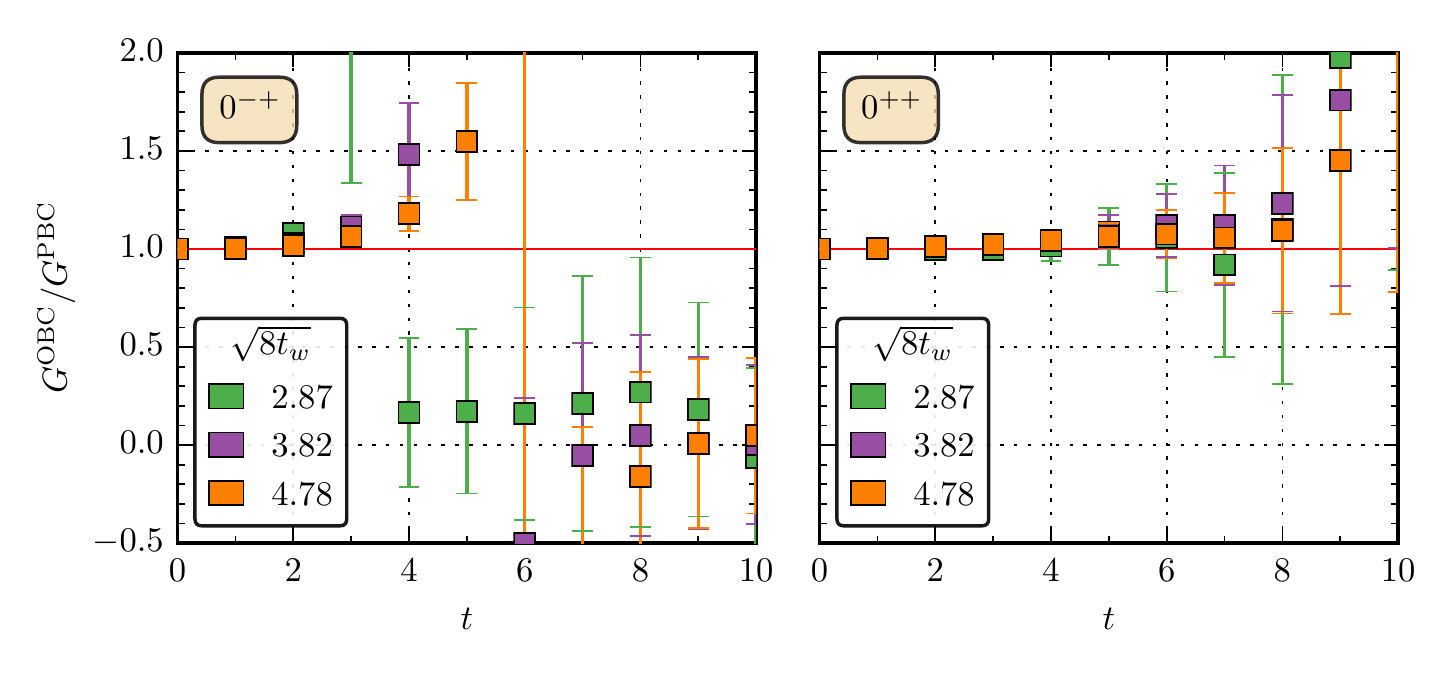}
	\caption{Ratio of correlators of gluonic operators on PBC and on OBC for (left) plaquette and (right) topological charge density, at three values of the flow time $t_w$.} 
	\label{fig:glue-corr-comp}
\end{figure}

We  smooth our configurations using the Wilson Flow, for various values of the flow time $t_w$. Our results are reported in Fig.~\ref{fig:glue-corr-comp}. As expected, the signal disappears into noise after few lattice spacings, but general trends can already be noticed. In particular, while the plaquette-plaquette correlator is compatible with one in the region in which it is determined with sufficient accuracy, it is possible to notice a tendency of the $q(t)$ correlator to deviate from one, signalling the presence of a lighter mode in OBC. Pending more detailed investigations, these observations seem to indicate that the lack of ergodicity with PBC does not significantly affect scalar glueball masses while at least the lighter pseudo-scalar glueballs are affected.

While the data are too noisy in the pseudoscalar channel, for the scalar one we attempted to extract the ground-state glueball mass, obtaining a result  of $a\,m_{0^{++}}=0.613(26)$. This is compatible with the same quantity determined in~\cite{Lucini:2004my}  for $SU(6)$ and $SU(8)$ at a comparable lattice spacing. 


\section{Conclusions}
The effects of the large-$N$ slowing down of topology have been investigated for SU(7) gauge theory by comparing the system on periodic and open temporal boundary lattices. 
The latter boundary conditions have proved to significantly decrease the correlation of the topological charge. However, our investigation has shown that in the bulk of the system, far from the boundary, the topological charge has a similar correlation time on the two systems. 
This confirms the expectations that when the system size is large enough the boundary does not influence the physics. 
The instanton size distribution shows that small instantons
are concentrated near the temporal boundaries of OBC lattices, with the instanton size distribution being the same in the bulk for OBC and PBC. 
This shows that the mechanism that changes the topological charge seems to be related to dislocation detaching from the boundary and growing into the bulk. While suppressed, this phenomenon seems to be frequent enough to determine a distribution of $Q$ that indicates good ergodicity through the topological sectors. 
The effects of topology on the glueball spectrum have also been investigated. The absence of any deviation of the plaquette-plaquette correlator with PBC from that obtained on OBC lattices seems to imply that a potential lack of ergodicity, as signalled by the distribution of $Q$, does not affect the scalar glueball tower. Conversely, the topological charge density correlator with OBC conditions deviates from that with PBC.
Future directions of this work include a comparison of the full glueball spectrum on the two setups using variational methods and Wilson flow techniques, in order to expose potential systematic effects in ground-state and excited masses in various channels. In addition, to get a better understanding of the dynamics of the topological modes, further investigations at larger $N$ and smaller lattice spacings should be performed.

\section*{Acknowledgements}
We thank M. Teper for discussions. AA is supported by the Academy of Finland (grant 1267286).  The work of BL was supported by STFC (grant ST/L000369/1). GB was supported by the DFG SFB/TR 55. 
Numerical computations were performed in part on the HPC Wales Sandybridge cluster at Swansea, supported by the ERDF through the WEFO (part of the Welsh Government).  The code used in this work is written using a GPU implementation of the QDP++ library~\cite{Winter:2014dka}.


\begin{thebibliography}{99}

\bibitem{obc}
M.~L\"uscher, 
PoS LATTICE2010  (2010) 015;
M.~L\"uscher, S.~Schaefer,
  JHEP 1107 (2011) 036;
M.~L\"uscher, S.~Schaefer, 
   Comput. Phys. Commun. 184 (2013) 519--528.

\bibitem{indobc}
A.~Chowdhury, A.~Harindranath, J.~Maiti, 
   JHEP 1406 (2014) 067;
%
A.~Chowdhury, A.~Harindranath, J.~Maiti, P.~Majumdar, 
  JHEP 1402 (2014) 045;
%
A.~Chowdhury, A.~Harindranath, J.~Maiti, 
   Phys. Rev. D91~(7) (2015) 074507.

\bibitem{obc-spectrum}
  M.~Bruno, P.~Korcyl, T.~Korzec, S.~Lottini and S.~Schaefer,
  PoS LATTICE {\bf 2014} (2014) 089;
%
  W.~Söldner [RQCD Collaboration],
  PoS LATTICE {\bf 2014} (2015) 099.
  
\bibitem{Bali:2013kia}
G.~S. Bali, F.~Bursa, L.~Castagnini, S.~Collins, L.~Del~Debbio, et~al., {Mesons
  in large-N QCD}, JHEP 1306 (2013) 071.

\bibitem{Teper:1985rb}
M.~Teper, {Instantons in the Quantized SU(2) Vacuum: A Lattice Monte Carlo
  Investigation}, Phys. Lett. B162 (1985) 357.

\bibitem{Luscher:2010iy}
M.~L\"uscher, {Properties and uses of the Wilson flow in lattice QCD}, JHEP 1008
  (2010) 071.

\bibitem{Bonati:2014tqa}
C.~Bonati, M.~D'Elia, {Comparison of the gradient flow with cooling in $SU(3)$
  pure gauge theory}, Phys. Rev. D89~(10) (2014) 105005.

\bibitem{Lucini:2004yh}
B.~Lucini, M.~Teper, U.~Wenger, {Topology of SU(N) gauge theories at T =~ 0 and
  T =~ T(c)}, Nucl. Phys. B715 (2005) 461--482.

\bibitem{Wolff:2003sm}
U.~Wolff, {Monte Carlo errors with less errors}, Comput. Phys. Commun. 156 (2004)
  143--153.

\bibitem{Schafer:1996wv}
T.~Schäfer, E.~V. Shuryak, {Instantons in QCD}, Rev. Mod. Phys. 70 (1998)
  323--426.
\bibitem{Smith:1998wt}
D.~A. Smith, M.~J. Teper, {Topological structure of the SU(3) vacuum}, Phys. 
  Rev. D58 (1998) 014505.
%
\bibitem{Lucini:2001ej}
B.~Lucini, M.~Teper, {SU(N) gauge theories in four-dimensions: Exploring the
  approach to N = infinity}, JHEP 06 (2001) 050.
\bibitem{Lucini:2004my}
B.~Lucini, M.~Teper, U.~Wenger, {Glueballs and k-strings in SU(N) gauge
  theories: Calculations with improved operators}, JHEP 0406 (2004) 012.
\bibitem{Winter:2014dka}
  F.~T.~Winter, M.~A.~Clark, R.~G.~Edwards and B.~Joó,
  A Framework for Lattice QCD Calculations on GPUs,
  doi:10.1109/IPDPS.2014.112.
 

\end{thebibliography}
\end{document}